\documentclass[preprint,prd,
amsmath,amssymb,amsthm,nofootinbib]{revtex4}
\usepackage[mathscr]{euscript}
\usepackage{bm}
\usepackage{textcomp}
\usepackage{graphicx}
\usepackage{subfigure}
\usepackage{overpic}
\usepackage{multirow}
\usepackage{floatrow}
\usepackage{float} 
\usepackage{amsmath,amssymb,amsthm}
\usepackage{cases}
\usepackage[colorlinks=true,linkcolor=red]{hyperref}
\newfloatcommand{capbtabbox}{table}[][\FBwidth]
\newcommand{\bea}{\begin{eqnarray}}
\newcommand{\eea}{\end{eqnarray}}
\newcommand{\beq}{\begin{equation}}
\newcommand{\eeq}{\end{equation}}

\def\/{\over}

\begin{document}

\title{Growth of  power spectrum due to decrease of sound speed during inflation}

\author{Rongrong Zhai, Hongwei Yu
\footnote{hwyu@hunnu.edu.cn} 
 and  Puxun Wu
 \footnote{pxwu@hunnu.edu.cn}  
 }
\affiliation{Department of Physics and Synergetic Innovation Center for Quantum Effects and Applications, Hunan Normal University, Changsha, Hunan 410081, China 
}

\begin{abstract}
We study   the amplification of the curvature perturbations due to a small sound speed  and find that its origin is different completely  from that due to the ultraslow-roll inflation. This is because when the sound speed is very small the enhancement of the power spectrum comes from the fact that the curvature perturbations at the scales smaller than the cosmic microwave background (CMB) scale becomes scale-variant, rather than  growing that leads to the amplification of the curvature perturbations during the ultraslow-roll inflation.  At  large scales the power spectrum of the curvature perturbations remains to be scale invariant, which is consistent with the CMB observations, and then it will have a transient $k^2$ growth and finally approach a $k^4$ growth  as the  scale becomes smaller and smaller. Thus  the power spectrum can be enhanced to generate a sizable amount of primordial black holes. Furthermore,  when  the high order correction in the dispersion relation of the curvature perturbations is considered the growth of the power spectrum of the curvature perturbations has the same origin as that in the case without this correction. 

\end{abstract}


\maketitle
\section{Introduction}
\label{sec_in}

Black holes, as one of the most mysterious objects in nature, can be formed from the supernova explosion of massive stars. It can also be generated in the very early Universe from the matter collapse resulting from  large density perturbations and such a black hole is  dubbed a primordial black hole (PBH)~\cite{Hawking1971,Carr1974,Carr1975}. These density perturbations originate from the curvature fluctuations during inflation~\cite{Starobinsky1980,Guth1980,Linde1982, Albrecht1982}, which are stretched to outside the Hubble horizon by the exponential cosmic expansion with the amplitudes frozen at certain values.  When these superhorizon fluctuations reenter the Hubble horizon during the radiation- or matter-dominated era, they will  lead to the formation of  large scale cosmic structures  and at the same time may also result in the PBH production.

To generate a sizable amount of PBHs, it is required that the amplitude $\mathcal{P_R}$ of the power spectrum of primordial curvature perturbations reaches the $\mathcal{O} (10^{-2})$ order, 
as has been demonstrated by  extensive  studies carried out in,  for example, Refs.~\cite{S. Chongchitnan, K. Kohri, Byrnes2019, E. Bugaev}. However, the cosmic microwave background (CMB) observations have limited the  primordial curvature perturbations to be about $10^{-5}$ with the amplitude of the corresponding power spectrum  being about $10^{-9}$ at the CMB scale~\cite{Aghanim2020}. Thus, to be consistent with the CMB observations and at the same time to generate a sizable amount of PBHs,  the primordial curvature perturbations can only  be enhanced at scales smaller than the CMB scale  during inflation. Since $\mathcal{P_R}\propto \frac{1}{c_s\epsilon}$, where $\epsilon=-\frac{\dot{H}}{H^2}$ is the slow-roll parameter, with $H$ being the Hubble parameter and $\dot{H}=\mathrm{d}H/\mathrm{d} t$,  and $c_s$ is the sound speed of the curvature perturbations, the enhancement of the curvature perturbations during inflation can be achieved by reducing the rolling speed of the inflaton to obtain a phase of  ultraslow-roll inflation~\cite{Germani2017,Motohashi2017,Ezquiaga2017,H. Di2018,Ballesteros2018,Dalianis2019,Gao2018,Drees2021,C.Fu2020, Xu2020, Lin2020, Dalianis2021, Yi2021,Gao2021, Yi2021b, TGao2021, Solbi2021, Gao2021b, Solbi2021b, Zheng2021,Teimoori2021a, Cai2021, Wang2021, Fuchengjie2019,fuchengjie2020,Dalianis(2020),Teimoori2021, Heydari2022, Heydari2022b,Bhaumik2020,Karam2022} or  the sound speed of the curvature perturbations~\cite{Ballesteros2022,Gorji2022,Ballesteros2019,Romano2016,Romano3,Ozsoy2018}.  

Slowing down the inflaton can be naturally realized by flattening the inflationary potential as well as by increasing the friction~\cite{Fuchengjie2019, Germani2011_1,Germani2011_2,Tsujikawa2012} or introducing a noncanonical kinetic term~\cite{Lin2020}. When the cosmic evolution changes from the slow-roll inflation to the ultraslow-roll one, the slow-roll parameter $\eta$, which is defined as $\eta=\frac{\dot{\epsilon}}{\epsilon H}$, will change from $\sim0$ to $-6$.  Solving the equation of motion for  the curvature perturbations, one can find that   the solution for  modes outside the Hubble horizon during the slow-roll inflation consists of a constant term and a decaying one,  which leads to a nearly scale-invariant spectrum. Assuming that the Universe  transitions suddenly  at  a time  from the slow-roll inflation  to  the ultraslow-roll one, we can match the solution of the curvature perturbations and its first derivative at this time by the Israel junction conditions~\cite{Israel1966,Deruelle1995}. After this transition, the term, which is decaying with time before the transition, becomes ``growing".  At first, the solution is dominated by the constant term and the power spectrum of the curvature perturbations remains to be scale invariant [$\mathcal{P_R}(k)\propto k^0$]. If the ultraslow-roll inflation lasts sufficiently long, the growing term will become dominant, which results in a $k^4$-dependent power spectrum~\cite{Garrilho2019,Byrnes2019,Liu2020}.  Because the constant and growing terms cancel each other, the power spectrum has a dip   preceding the $k^4$ dependence. If an $\eta=-1$  middle phase  is added between the slow- and ultraslow-roll inflations, it has been found that the power spectrum has a $k^5 (\log k)^2$ growth~\cite{Garrilho2019,Cole2022}.  

Similar to the case of a tiny $\epsilon$, a smaller value of the sound speed can also lead to an amplification of the primordial curvature perturbations. For a scalar field with a standard kinetic term, the sound speed equals  the light speed ($c_s=1$).  A varying $c_s$ can be realized  in many inflationary models including the inflation with a noncanonical kinetic term~\cite{Armendariz-Picon1999,Garriga1999,Ballesteros2019,A. Y. Kamenshchik},  the Dirac-Born-Infeld inflation~\cite{Kamenshchik2022},  the multifield inflationary model~\cite{Romano1,Romano2}, the inflationary models in modified gravity theories~\cite{CGermaniprl,S.Tsujikawa}, and so on. 
 Using the effective field theory of inflation, the generation of PBHs due to a small $c_s$ has been studied in~\cite{Ballesteros2022,Gorji2022,Ballesteros2019, Romano3}. However, whether the growth of the power spectrum due to  decreasing sound speed is similar to that due to ultraslow-roll inflation and whether this growth results from the appearance of a growing term in the solution of  the curvature perturbations  remains  unclear.   Recently, it was pointed out that when the sound speed is very small  the high order  $k^4$ term  in the dispersion relation will become dominant~\cite{Ballesteros2022,Gorji2022}. Although this high order term leads to a new solution of the curvature perturbations, the power spectrum is enhanced by  small $c_s$ and it has a $k^4$ growth, which is the same as that in the case of the ultraslow-roll inflation. Can both $k^4$ growths  of the power spectrum have the same origin?  The purpose of the present paper is to answer the above mentioned questions by studying in detail  the growth of the power spectrum of the curvature perturbations due to the decrease of the sound speed.  

The paper is organized   as follows: In Sec.~\ref{sec2}, we investigate the growth of the power spectrum due to a sudden decrease of the sound speed.    In Sec.~\ref{sec3},   the effect of the high order  $k^4$ term in the dispersion relation is analyzed.   Finally, we give our conclusions in Sec.~\ref{conclusion}. Throughout this paper, we set $c=\hbar=M_\mathrm{Pl}=1$.

\section{the growth of power spectrum from a small sound speed}
\label{sec2}
To study the growth of the power spectrum of the curvature perturbations due to a sudden decrease of the sound speed, we need to consider the Sasaki-Mukhanov equation
\bea\label{vk1}
	v''_k+\left(c_s^2k^2-\frac{z''}{z}\right)v_k=0
 \eea
in  Fourier space,  where a dash denotes  a differentiation with respect to the conformal time $\tau$,  $v_k=z \mathcal{R}_k$, $\mathcal{R}$ is the  curvature perturbation, and $z$ is defined to be 
 \bea\label{z}
 z^2 \equiv \frac{2a^2 \epsilon}{c_s^2}
 \eea
with $a$ being the cosmic scale factor. From Eq.~(\ref{z}), one can obtain that 
\bea\label{z2}
\frac{z''}{z}=(aH)^2\left(2-\epsilon+\frac{3}{2}\eta-3s+s^2+s\epsilon-s \eta+ \frac{1}{4}\eta^2-\frac{1}{2}\eta\epsilon\right) \; ,
\eea
where $s=\frac{\dot{c}_s}{c_s H}$.  Equation~(\ref{z2}) can be reexpressed to be
\bea
\frac{z''}{z}=\frac{\nu^2-1/4}{(-\tau)^2} \; ,
\eea
where $aH=-\frac{1}{\tau}$ has been used, and $\nu\simeq \frac{3}{2}+\frac{1}{3}\epsilon+\frac{1}{2}\eta-s$. 
If $c_s$ is a constant, the general solution of Eq.~(\ref{vk1}) has the form
\bea\label{vk2}
v_k (\tau)=A \sqrt{-  \tau } H_\nu^{(1)}(-c_s k \tau )+B \sqrt{-  \tau } H_\nu^{(2)}(-c_s k \tau )\;.
\eea
Here $A$ and $B$ are two constants. Thus, the solution of the curvature perturbations can be obtained through the relation $\mathcal{R}_k=\frac{ v_k}{z}$.

Now we consider that  during the slow-roll inflation the sound speed $c_s$ will change suddenly from $1$ to a very small value denoted by $1/A_s$ at conformal  time $\tau_1$, where $A_s$ is a constant and $A_s\gg 1$.  Since $\nu \simeq-3/2$ for the slow-roll inflation with a constant sound speed, we find that when $c_s=1$ the solution of the curvature perturbations has the form 
\bea\label{R1}
\mathcal{R}^{(1)}_k(\tau)= i\frac{H}{2\sqrt{\epsilon k^3}}e^{-ik\tau}(1+ik\tau)
\eea
after choosing the adiabatic Bunch-Davis vacuum condition. 

For $c_s=1/A_s$, from Eq.~(\ref{vk2}), we obtain the general solution of the curvature perturbations 
\bea\label{R2}
\mathcal{R}^{(2)}_k (\tau)=\frac{H}{2\sqrt{\pi A_s\epsilon k^3}}\left[ A_2 e^{-\frac{ik\tau}{A_s}}(A_s+ik\tau)-B_2 e^{\frac{ik\tau}{A_s}}(A_s-ik\tau) \right]\; ,
\eea
where $A_2$ and $B_2$ are two constants. Matching $\mathcal{R}^{(1)}_k$ and $\mathcal{R}^{(2)}_k$ at $\tau=\tau_1$ by using the conditions $\mathcal{R}^{(1)}_k (\tau_1)=\mathcal{R}^{(2)}_k (\tau_1)$ and $\mathcal{R}'^{(1)}_k(\tau_1)=\mathcal{R}'^{(2)}_k (\tau_1)$, one can obtain that 
\bea
A_2 &=& \frac{\sqrt{A_s\pi}}{2k} (1+A_s)(-1+ik+A_s)e^{ik \big(1-\frac{1}{A_s} \big )} , \\
B_2 &=& \frac{\sqrt{A_s\pi}}{2k} (-1+A_s)(1-ik+A_s)e^{ik\big(1+\frac{1}{A_s}\big)} ,
\eea
where we have set $\tau_1=-1$, which simplifies the expressions but does not change  any physical result. Substituting $A_2$ and $B_2$ into Eq.~(\ref{R2}) gives the expression of the curvature perturbations during the phase of $c_s=1/A_s$.
 Expanding this expression in the infrared limit 
 ($c_s k \tau\rightarrow 0$), 
 we arrive at
	\bea\label{R2-t}
	\mathcal{R}^{(2)}_k (\tau) \simeq \frac{i H A_s e^{ik}} {2\sqrt{\epsilon k^3}} C_2 -\dfrac{i H A_s e^{ik}}{2\sqrt{\epsilon k^5}}D_2-\frac{iHe^{ik} } {4A_s\sqrt{\epsilon k }} (-\tau)^2 D_2+\cdots \; .
	\eea
	where
	\bea \label{CD}
	C_2&=& A_s \cos\left(\frac{k}{A_s}\right)-i \sin\left(\frac{k}{A_s}\right) \nonumber\\
	D_2&=& \left(A_s^2-1\right)\sin\left(\frac{k}{A_s}\right) .
	\eea
For modes that are superhorizon at $\tau_1$, we can further expand Eq.~(\ref{CD}) at $k\tau_1=-k\rightarrow0$ and then obtain
\bea\label{R2-t1}
\mathcal{R}^{(2)}_k (\tau) \simeq \frac{iH e^{ik}} {2\sqrt{\epsilon k^3}} +\dfrac{He^{ik}}{2\sqrt{\epsilon k}}-\frac{iH(2A_s^2+1)e^{ik} k^{1/2}} {12A_s^2\sqrt{\epsilon }}-\frac{iH(A_s^2-1)e^{ik} k^{1/2}} {4A_s^2\sqrt{\epsilon}} (-\tau)^2+\cdots \; .
\eea 
It is easy to see that the leading part is  constant independent of $\tau$, which contains three
 different $k$-dependent terms and the subleading part decays with time since $|\tau|$ decreases during inflation.  These characters are different from  that in  the case of the transition from the slow-roll inflation to the ultraslow-roll one, 
where there is a growing part and the constant part only has a $k^{-3/2}$-dependent term~\cite{Ballesteros2022}.
 
\begin{figure}[H]
	\centering
	\includegraphics[width=0.61\linewidth]{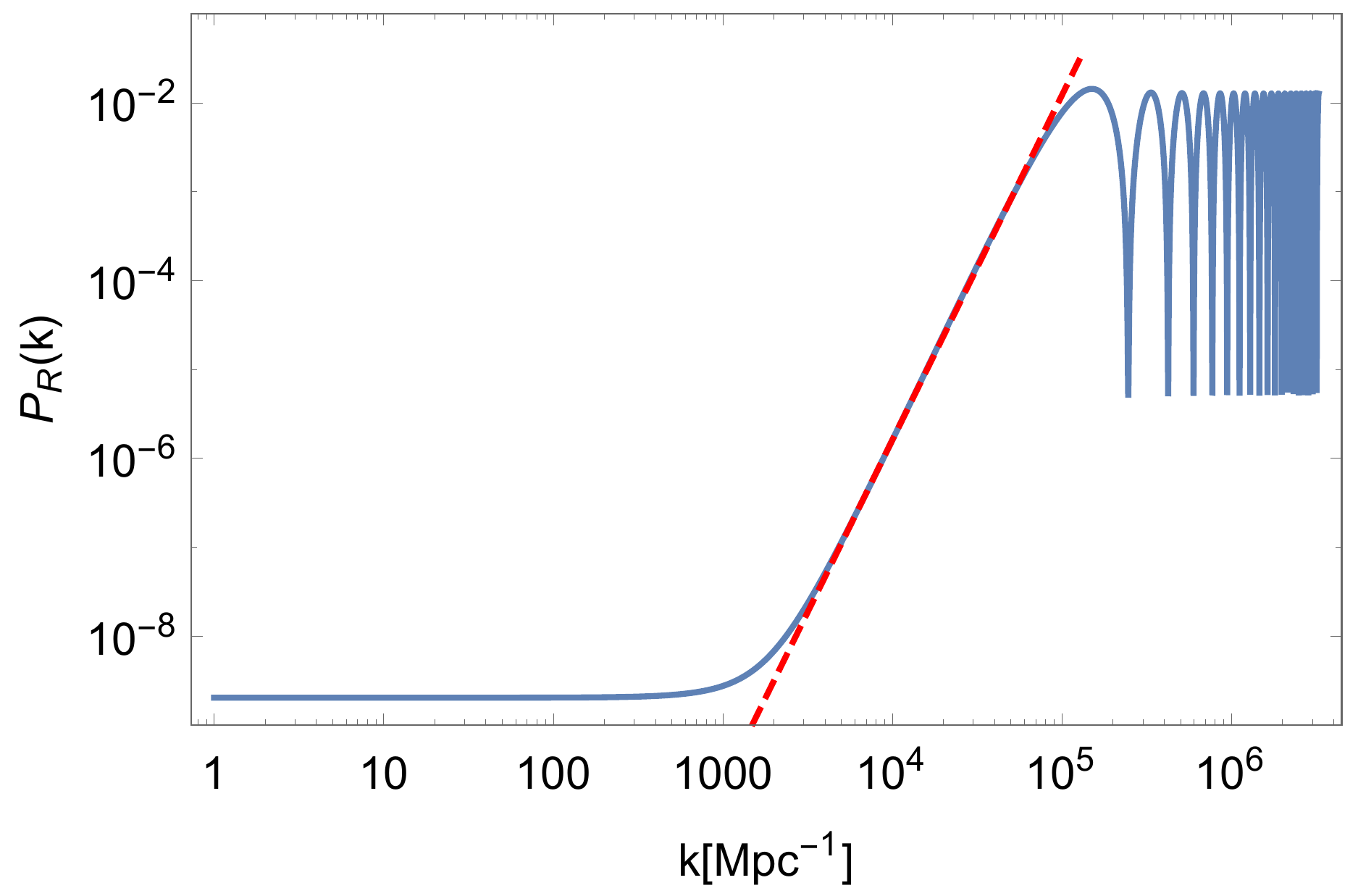}
	\caption{\label{fig1}  The evolution of power spectrum $\mathcal{P_R}$ as functions of wave number $k$ with $A_s=50$.
		The dashed red line indicates the $k^4$ slope.   }
\end{figure}

 Equation~(\ref{R2-t1}) gives that the  power spectrum of the curvature perturbations has the form 
\bea\label{Prt1}
\mathcal {P}_{\mathcal{R}^{(2)}_k} = \frac{k^3}{2\pi^2}|\mathcal{R}^{(2)}_k|^2\simeq \frac{H^2}{8\pi^2 \epsilon}+\frac{(A_s^2-1)H^2}{24 \pi^2  A_s^2 \epsilon} k^2  +\frac{(2A_s^2+1)^2 H^2}{2\left(12\pi A_s^2\right)^2 \epsilon} k^4 \;.
\eea
Apparently, the power spectrum of the curvature perturbations consists of a $k$-independent term and two $k$-dependent ones. At the CMB scale, the first term dominates, which leads to a scale-invariant spectrum consistent with the CMB observations. Going to the scales that are smaller than the CMB scale, the second term begins to play a role. The power spectrum becomes scale dependent and has a short era with a $k^2$ growth. Then, the third term finally dominates and the power spectrum displays a $k^4$ growth. These results are shown clearly in Fig~\ref{fig1}. This figure indicates that there is no dip in the power spectrum, although the dip appears in the power spectrum in the ultraslow-roll inflation. 
This  is because the second term in Eq.~(\ref{Prt1}) does not cancel the first term.  From Fig.~\ref{fig1}, one can see that the power spectrum oscillates at the small scales. The reason is that although these small scales are superhorizon at $\tau$, they are sub-horizon at $\tau_1$. For subhorizon scales at $\tau_1$, Eq.~(\ref{CD}) shows clearly that $C_2$ and $D_2$ are  oscillating functions, which results in an oscillating power spectrum. 
 
\section{The effect of higher order correction in dispersion relation}
\label{sec3}
When $c_s$ is very small, the high order correction in the dispersion relation of the curvature perturbations needs to be considered~\cite{Ballesteros2022,Gorji2022,Ballesteros2019}. When the $k^4$ corrected term in the dispersion relation is included, the numerical calculation shows that the power spectrum has a $k^4$ growth~\cite{Ballesteros2022,Gorji2022}. In this section, we discuss the origin of this growth.  We find that $v_k$ satisfies the following  equation
\bea\label{k4}
v_k''+\left[ c_s^2 k^2+ \alpha k^4(- \tau)^2-\frac{\nu^2-1/4}{(-\tau)^2}\right]v_k=0 \;.
\eea 
where $\alpha$ is a constant. When $c_s=1$, this $k^4$ corrected term can be neglected and $v_k$ has a general solution shown in Eq.~(\ref{vk2}). Thus, choosing the adiabatic Bunch-Davis vacuum condition the solution of $\mathcal{R}_k^{(1)}$ is given in Eq.~(\ref{R1}).

For a very small $c_s$, which implies a very large $A_s$ after setting $c_s=\frac{1}{A_s}$, the  $k^4$ corrected term  in Eq.~(\ref{k4})  will dominate over  $c_s^2 k^2$ and the latter can be neglected. Solving Eq.~(\ref{k4}) and using $\mathcal{R}_k=\frac{v_k}{z}$ and $\nu=-3/2$, we find that the curvature perturbation  has the general solution 
\bea
\mathcal{R}^{(2)}_k(\tau)=-\frac{\alpha^{1/8}H}{2A_s \sqrt{\epsilon}} k^{1/2} (-\tau)^{3/2}\left[A_3 H_{3/4}^{(1)}\left (\frac{1}{2} \sqrt{\alpha} k^2 \tau^2 \right)+ B_3 H_{3/4}^{(2)}\left (\frac{1}{2} \sqrt{\alpha} k^2 \tau^2 \right)\right] \;,
\eea
where $A_3$ and $B_3$ are two constants. Using the matching condition: $\mathcal{R}^{(1)}_k(\tau_1)=\mathcal{R}^{(2)}_k(\tau_1)$ and $\mathcal{R}'^{(1)}_k(\tau_1)=\mathcal{R}'^{(2)}_k(\tau_1)$ at time $\tau_1$ which is set to be $-1$, one can obtain that
\begin{align}  
	A_3 &= \frac{A_s e^{ik} ~C}{\alpha^{5/8}k^4~ E}\; ,\\
	B_3 &= \frac{A_s e^{ik}~D}{\alpha^{5/8}k^4~E}\; ,
\end{align}
where
\begin{align}  
	C =& i \sqrt{\alpha}(i+k) k^2 H_{-1/4}^{(2)} \left(\frac{1}{2}\sqrt{\alpha }k ^2  \right)-(3-3ik-2k^2)H_{3/4}^{(2)} \left(\frac{1}{2}\sqrt{\alpha }k^2  \right)\nonumber\\&+\sqrt{\alpha}(1-i k)k^2 H_{7/4}^{(2)} \left(\frac{1}{2}\sqrt{\alpha }k^2  \right),\nonumber\\
	D=& \sqrt{\alpha}(1-ik)k^2 H_{-1/4}^{(1)} \left(\frac{1}{2}\sqrt{\alpha }k ^2  \right)+(3-3ik-2k^2)H_{3/4}^{(1)} \left(\frac{1}{2}\sqrt{\alpha }k^2  \right)\nonumber\\&- \sqrt{\alpha}(1-i k)k^2 H_{7/4}^{(1)} \left(\frac{1}{2}\sqrt{\alpha }k^2  \right),\nonumber \\
	E=&H_{3/4}^{(2)} \left(\frac{1}{2}\sqrt{\alpha } k^2 \right)\left[H_{-1/4}^{(1)} \left(\frac{1}{2}\sqrt{\alpha }k^2  \right)-H_{7/4}^{(1)} \left(\frac{1}{2}\sqrt{\alpha }k^2\right)\right]\nonumber\\&+
	H_{3/4}^{(1)} \left(\frac{1}{2}\sqrt{\alpha } k^2 \right)\left[H_{7/4}^{(2)} \left(\frac{1}{2}\sqrt{\alpha } k^2 \right)-H_{-1/4}^{(2)} \left(\frac{1}{2}\sqrt{\alpha } k^2 \right)\right].
\end{align}
  In the infrared region, we have
	\bea\label{R2-2}
	\mathcal{R}^{(2)}_k (\tau) \simeq \frac{H e^{ik}} {2\sqrt{\epsilon k^3}} -\dfrac{i He^{ik}}{2\sqrt{\epsilon k}}-\frac{He^{ik} k^{1/2}} {6\sqrt{\epsilon }}+\frac{He^{ik} k^{1/2}} {6\sqrt{\epsilon }} (-\tau)^3+\cdots \; .
	\eea
This solution consists of three constant terms and the decaying terms, but there is no  growing term.
 This is similar to the case studied in the previous section.
 Then, we obtain the power spectrum during the small sound speed phase,
\bea\label{Pr2}
\mathcal {P}_{\mathcal{R}^{(2)}_k}= \frac{k^3}{2\pi^2}|\mathcal{R}^{(2)}_k|^2 \simeq \frac{H^2}{8\pi^2 \epsilon}+\frac{H^2}{24 \pi^2 \epsilon} k^2  +\frac{H^2}{72\pi^2 \epsilon} k^4 \;.
\eea
This  is the same as that  given in Eq.~(\ref{Prt1}) since $A_s\gg1$.  
 So, the high order correction in the dispersion relation has no  contribution to the growth of the power spectrum of the curvature  perturbations.

 \section{conclusion}
 \label{conclusion}
 A generation of abundant PBHs in the early Universe requires that the amplitude of the power spectrum of the primordial  curvature perturbations is enhanced by about 7 orders during inflation, which can be realized by the ultraslow-rolling of inflaton, a very small sound speed of the curvature perturbations, and even some other mechanisms. It has been found that in the  ultraslow-roll inflation   the amplification of the power spectrum  can be attributed to the appearance of a growing solution of the curvature perturbations. In this paper, we find that  the origin of the amplification of the curvature perturbations due to a small sound speed  is different completely from that in the case of the ultraslow-roll inflation. This is because when the sound speed is very small the enhancement of the power spectrum comes from the fact that the curvature perturbations at the scales smaller than the CMB scale becomes scale variant, rather than growing. The power spectrum of curvature perturbations remains to be scale invariant  at  large scales, and then it has a short time $k^2$ growth and finally approaches a $k^4$ growth  as the scale becomes smaller and smaller.    Thus, the curvature perturbations during inflation with a decrease of the sound speed can be consistent with the CMB observations and at the same time enhanced to generate a sizable amount of PBHs. Furthermore, we find that when  the high order correction in the dispersion relation of the curvature perturbations is considered, the growth of the power spectrum of the curvature perturbations has the same origin  as that  without this correction.

 \begin{acknowledgments}
We appreciate very much the insightful comments and helpful suggestions by the anonymous referee. This work is supported by the National Key Research and Development Program of China Grant No. 2020YFC2201502,  the National Natural Science Foundation of China under Grants No.  12075084, No. 11775077  and No. 11805063, and  the Science and Technology Innovation Plan of Hunan Province under Grant No. 2017XK2019.
\end{acknowledgments}

\end{document}